\documentclass[prl,aps,amssymb,twocolumn,showpacs,superscriptaddress]{revtex4}
\usepackage{graphicx}

\begin{document}

\title{The Noise Susceptibility of a Photo-excited Coherent Conductor}

\author{J.~Gabelli}
\author{B.~Reulet}
\affiliation{Laboratoire de Physique des Solides, UMR8502, b\^atiment 510, Universit\'e Paris-Sud, 91405 ORSAY Cedex, France}

\date{\today}
\begin{abstract}

We report the theory of the \emph{dynamical response of current fluctuations} of a photo-excited conductor. We have performed the calculation for a coherent conductor described by arbitrary energy-dependent transmissions and for arbitrary frequencies. We consider two experimental setups that correspond to different ways of symmetrizing the current operators, leading to different predictions. Our results are in very good, quantitative agreement with a recent measurement. We demonstrate that the dynamical response of noise that we have calculated is the key concept that relates Dynamical Coulomb Blockade, i.e. the electron-electron correction to the conductance due to the presence of an external impedance, to quantum noise.
\end{abstract}
 \maketitle

Measuring the dc current $I$ through a conductor while applying a dc voltage $V$ (which gives the differential conductance $dI/dV$) provides information only about dc properties of conduction. To learn more, several directions can be taken. First, one can probe the \emph{dynamics of the charge transfer} through the sample by measuring the dynamical conductance $G(\omega_0)$ at finite frequency $\omega_0$. This is achieved by applying dc plus small ac voltage $V(t)=V+\delta V\cos\omega_0 t$ and measuring the current at the same frequency. Since the current may have both in-phase and out-of-phase components, $G(\omega_0)$ is in general a complex quantity. If $\delta V$ is large one may even enter a non-linear regime, in which current at harmonic frequencies $p\omega_0$ will be generated ($p$ is an integer). A second option is to look at current fluctuations instead of average current. Current fluctuations are characterized by their spectral density $S_2(\omega)$ measured at frequency $\omega$: $S_2(\omega)=\langle i(\omega)i(-\omega)\rangle$. Here $i(\omega)$ is the Fourier component of the fluctuating current at frequency $\omega$; the brackets $\langle.\rangle$ denote time averaging. $S_2$, often simply called "noise", can be studied as a function of voltage and frequency (for a review on noise, see \cite{BuBlan}). It provides information about the \emph{statistics of the charge transfer}.

Again, to learn more one is naturally drawn to consider noise under dc+ac bias. Photo-assisted noise corresponds to the noise $S_2(\omega)$ in the presence of an excitation at frequency $\omega_0$, obtained by time averaging the square of the current filtered around $\omega$, as in \cite{PAT_Rob} for $\omega=0$ and in \cite{JGBRmanip} for $\omega\sim\omega_0$. The equivalent of the dynamical response of current at arbitrary frequencies $p\omega_0$ with $p\neq0$ has however never been calculated for noise ($p=0$ is the usual photo-assisted noise \cite{BuBlan}). It is the goal of this paper to provide such a calculation, which involves the correlator $\langle i(\omega) i(p\omega_0-\omega)\rangle$. For a small ac voltage, we define the noise susceptibility which is to the noise what the ac conductance is to the average current. It is different from the adiabatic response obtained by taking the derivative of $S_2$ vs. $V$, just as the dynamical conductance $G(\omega_0)$ differs from its dc value. At low excitation frequency $\omega_0$, one can reformulate the noise susceptibility in other terms, as follows. The ac voltage makes the Joule power dissipated in the sample to oscillate at $\omega_0$ by a quantity $\delta P_J^{\omega_0}=2GV\delta V$ for $G$ real. From the noise temperature $T_N=S_2/(4k_BG)$ (here we suppose $S_2$ and $G$ independent of $\omega$ for the sake of simplicity), the \emph{noise thermal impedance}  (NTI) is defined as the ratio of the oscillating noise temperature $\delta T_N^{\omega_0}$ to the oscillating Joule power \cite{NTI}: $\mathcal{R}(\omega_0)=\delta T_N^{\omega_0}/\delta P_J^{\omega_0}$. It is well defined only if the noise temperature can be measured in a time much shorter than the inverse excitation frequency, i.e. $\omega\gg\omega_0$. The NTI is simply proportional to the small $\omega_0$ limit of the noise susceptibility that we calculate for arbitraries $\omega$ and $\omega_0$.  The noise susceptibility measures the \emph{dynamics of (charge neutral) electron-hole excitations}, which reveals information that dynamics or statistics of charge transfer (given respectively by the ac conductance $G(\omega)$ and by $S_2(\omega)$) do not provide because of strong screening occurring in good conductors.

In the following we calculate the correlator $\langle i(\omega) i(p\omega_0-\omega)\rangle$ for a quantum conductor using the Landauer-B\"uttiker formalism, along the lines of ref. \cite{BuBlan}. In order to make a link with experiments, one has to symmetrize operators properly. We show two different experimental setups that correspond to different symmetrization rules, resulting in different shapes for the voltage dependence of the signal. Our calculation reproduces very well a recent measurement performed on a tunnel junction in the regime $\omega\sim\omega_0$ \cite{JGBRmanip}. We then show that the noise susceptibility, i.e. the linear response of noise to a small ac voltage, is the central concept that relates Dynamical Coulomb blockade (DCB) to quantum noise. We indeed demonstrate that usual expressions for DCB can be rewritten in terms of noise susceptibility.

We consider a single quantum channel described by its transmission amplitude $t(E)$ and reflection amplitude $r(E)$. We will perform all the calculations for a single channel. The formulas for the many channels case are obtained by simply adding the contributions of all the channels. The  current operator at frequency $\omega$ is given by \cite{BuBlan}:
$$
\begin{array}{rl}
I(\omega)=&  \frac he \int dE [
(1-r^*(E)r(E+\hbar\omega))a^+_L(E)a_L(E+\hbar\omega)\\
&-r^*(E)t(E+\hbar\omega)a^+_L(E)a_R(E+\hbar\omega) \\
& -t^*(E)r(E+\hbar\omega)a^+_R(E)a_L(E+\hbar\omega)\\
&-t^*(E)t(E+\hbar\omega)a^+_R(E)a_R(E+\hbar\omega) ]
\end{array}
$$
$a_{L,R}(E)$ is the annihilation operator of a quasiparticle of energy $E$ in the left (right) reservoir.
 The effect of the dc+ac bias $V(t)=V+\delta V\cos\omega_0t$ is to shift adiabatically the Fermi sea of the left reservoir, which is accounted for by replacing $a_L$ by \cite{Tien}:
\begin{equation}
a_L(E)\rightarrow \tilde a_L(E)=\sum_{n=-\infty}^\infty J_n(z)
a_L(E-eV-n\hbar\omega_0)
\end{equation}
where $z=e\delta V/(\hbar\omega_0)$ and $J_n$ is the $n^{th}$ Bessel's function of the first kind. We neglect electron-electron interactions, which is valid provided that $\omega_0$ is smaller than frequencies associated with electrodynamics, i.e. the plasma frequency or inverse of RC-like time \cite{BuAC}. The statistical averages are given by:
$$
\begin{array}{rl}
 \left<\tilde a^+_L(E)\tilde a_L(E+\hbar\omega)\right> &= \sum_{p} f_L^{(p)}(E)\delta(\hbar\omega-p\hbar\omega_0)\\
\left<a^+_R(E) a_R(E+\hbar\omega)\right> &= f_R(E) \delta(\hbar\omega)\\
\end{array}
$$
with $f_L^{(p)}(E)= \sum_n J_n(z)J_{n+p}(z)f_L(E-n\hbar\omega_0)$ an effective energy distribution function and $f_L$, $f_R$ the Fermi distributions in the left and right contacts.
We now consider how the noise is modulated by the ac excitation at frequency $\omega_0$. The effect of the ac
excitation is to induce correlations between Fourier components of the current separated by $p\omega_0$ where $p$ is an integer. Such correlations are calculated by the mean of the following quantity:
\begin{equation}
X_+^{(p)}(\omega_0,\omega)=\left<I(\omega)I(p\omega_0-\omega)\right>-\left<I(\omega)\right>\left<I(p\omega_0-\omega)\right>
\end{equation}
which, unlike noise or photo-assisted noise, is complex for $p\neq0$. The case $p=0$ corresponds to the photo-assisted noise. For $p\neq 0$, $X_+^{(\pm p)}$ measures how noise measured at $\omega$ oscillates at frequency $p\omega_0$. In the case of a small excitation, only $X_+^{(\pm1)}$ remains, which can be interpreted as a susceptibility of noise to an external oscillating voltage $X_+^{(\pm1)}(\omega) \propto \delta S_2(\omega)/\delta V$.
 In the case of a slow excitation, the noise follows adiabatically the voltage, and one expects $X_+^{(\pm1)}(\omega) \sim \delta V(dS_2(\omega)/dV)$. We decompose $X_+^{(p)}$ into:
\begin{equation}
\begin{array}{c}
  X_+^{(p)}(\omega_0,\omega)=  \frac{e^2}{h^2} \int dE
[x_{LR}(E)+x_{RL}(E)]
\vspace{2mm}\\
\begin{array}{rl}
 x_{LR} = &  |t(E+\hbar\omega)|^2r^*(E)r(E+p\hbar\omega_0)\\
 & \hspace{2cm} f_L^{(p)}(E)(1-f_R(E+\hbar\omega)) \\
 x_{RL} = &
  |t(E)|^2r(E+\hbar\omega)r^*(E+\hbar\omega-p\hbar\omega_0) \\
&  \hspace{1.5cm} f_R(E)(\delta_{p,0}-f_L^{(-p)}(E+\hbar\omega))
\end{array}
\end{array}
\label{eqgeneX}
\end{equation}
We remark that $X_+^{(p)}$ is real for energy-independent transmissions. In this case (which corresponds to zero dwell time for the electrons), the noise responds instantaneously to the excitation, but as we will see, not adiabatically. One now has to take care of the proper symmetrization of $X_+^{(p)}$ in order to calculate what corresponds to an actual measurement. It appears that the quantity one has to calculate depends on the experimental setup. We will consider here two different setups.

\begin{figure}
\includegraphics[width= 0.99\columnwidth]{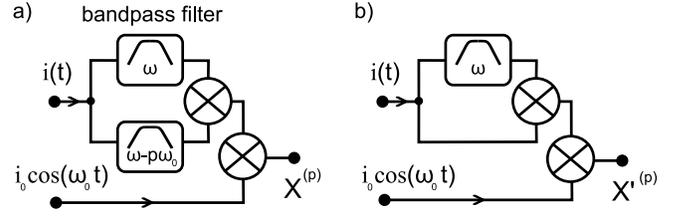}
\caption{ (a) Detection scheme for the two filters correlator $X(\omega_0,\omega)$. (b) Detection scheme for the one filter correlator $X'(\omega_0,\omega)$. The symbol $\otimes$ represents a multiplier, which output is the product of its two inputs.}
\label{Fig:filters}
\end{figure}

\emph{Detection with two filters at $\omega$ and $\omega-p\omega_0$ ---}
We first consider the setup of Fig. \ref{Fig:filters}(a). The current $i(t)$ is split into two paths where it is filtered around different frequencies: in the upper arm, it is filtered around frequency $\omega>0$, to give  $i_\omega(t)=i(\omega)e^{i\omega t}+i(-\omega)e^{-i\omega t}$; in the lower arm it is filtered around frequency $|\omega-p\omega_0|$ ($p>0$, $\omega_0>0$). The response of the noise in-phase with the excitation is obtained by multiplying this product by $\cos p\omega_0 t$ with $\omega_0>0$ (and the out-of-phase by $\sin p\omega_0 t$) and then taking the dc part by time averaging. This gives the quantity $X^{(p)}(\omega_0,\omega)= \left<i_\omega(t)i_{p\omega_0-\omega}(t)\cos p\omega_0 t\right>$:
\begin{equation}
\begin{array}{rl}
X^{(p)}(\omega_0,\omega)= & \left[ \left<i(\omega)i(p\omega_0-\omega)\right> \right. \\
& + \left. \left<i(-\omega)i(-p\omega_0+\omega)\right> \right]/2
\end{array}
\end{equation}

It is remarkable that $X^{(p)}(\omega_0,\omega)$ is not invariant upon the transformations $\omega\rightarrow\-\omega$ or $\omega_0\rightarrow-\omega_0$ taken separately, but is invariant if we change the sign of both frequencies. It is also invariant upon the transformation $\omega\rightarrow p\omega_0-\omega$. Each product has now to be symmetrized to be calculated by operators, since we are only considering a detection by linear elements and not by photon counters (which can only absorb photons) :
$$
\left<i(\omega_1)i(\omega_2)\right>\rightarrow
(\left<I(\omega_1)I(\omega_2)\right> +\left<I(\omega_2)I(\omega_1)\right>)/2
$$
Here the brackets around $i$ mean time average and around $I$ quantum mechanical average. Note that for the usual noise $\langle i(\omega)i(-\omega)\rangle$ this symmetrization is equivalent to the one on sign reversal of the frequency. This is not the case here: the symmetrization rule on frequencies (and thus on operators in the final expression) depends on the detection scheme.  Using the general property: $X_+^{(p)}(\omega_0,\omega)^*=X_+^{(-p)}(\omega_0,\omega-p\omega_0)$ one can simplify the expression of $X^{(p)}$. Moreover, if $X_+^{(p)}$ is real, as in the case of energy-independent transmission, the noise responds in phase with the excitation and $X^{(p)}$ simplifies:

\begin{equation}
X^{(p)}(\omega_0,\omega) = \left[ X_+^{(p)}(\omega_0,\omega)+X_+^{(-p)}(\omega_0,-\omega) \right] /2
\end{equation}
From Eq. (\ref{eqgeneX}) we find for the multi-channel case:
\begin{equation}
\begin{array}{l}
X^{(p)}(\omega_0,\omega)  =  (F/2) \sum_n J_n(z)J_{n+p}(z)\times  \\
  \hspace{1cm} \left[ S_2^0(\omega_+ +n\omega_0)
   +(-1)^p S_2^0(\omega_- +n\omega_0) \right]  \\
\end{array}
\label{eq:X}
\end{equation}
with $\omega_\pm=\omega\pm eV/\hbar$ and $S_2^0(\omega)=2G\hbar\omega\coth[\hbar\omega/(2k_BT)]$ the equilibrium Johnson-Nyquist noise at frequency $\omega$. $G$ is the conductance, supposed real and frequency-independent for the sake of simplicity; $F$ is the Fano factor. In the limit of small ac excitation $e\, \delta V\ll\hbar\omega_0$,
only the correlators for $p=\pm1$ contain terms linear in $\delta V$. We define the noise susceptibility by:
\begin{equation}
 \chi_{\omega_0}(\omega)=\lim_{\delta V \rightarrow 0}
\frac{X^{(1)}(\omega_0,\omega)}{\delta V}
\end{equation}
In the many channels case, the noise susceptibility is:
\begin{equation}
\begin{array}{rl}
 \chi_{\omega_0}(\omega)= & (F/2) e/(\hbar\omega_0) \left[
S_2^0(\omega_+)-S_2^0(\omega_-) \right.\\
& + \left. S_2^0(\omega_--\omega_0)-S_2^0(\omega_+-\omega_0) \right]
\end{array}
\label{eq:chi}
\end{equation}
and in the limit of a low frequency excitation $\omega_0\rightarrow 0$ it reduces to:
$\chi_0(\omega)=\frac12 \frac{dS_2(\omega)}{dV}$, where $S_2(V,\omega)$ is the usual noise given by \cite{BuBlan}:
\begin{equation}
S_2(V,\omega)=(F/2) \left[ S_2^0( \omega_+)+S_2^0 ( \omega_-)\right] +(1-F)S_2^0(\omega)
\label{eqSvsS0}
\end{equation}
The factor $1/2$ between $\chi_0(\omega)$ and $dS_2(\omega)/dV$ comes from the fact that the noise up-converted at frequency $\omega+\omega_0$ is not detected by this setup. If one slowly modulates the voltage and detect noise oscillations with a lockin technique, frequencies $\omega-\omega_0$ and $\omega+\omega_0$ are not separated (unlike with the present setup) and the full signal $dS_2/dV$ is recovered.

\emph{Detection with one filter at frequency $\omega$ ---}
Here we consider the detection scheme depicted on Fig.
\ref{Fig:filters}(b): in this setup the current filtered around frequency $\omega>0$,  $i_\omega(t)$, is multiplied by the unfiltered current $i(t)$. The result for the in-phase response $\langle i_\omega(t)i(t)\cos p\omega_0t\rangle$ is:
\begin{equation}
X'^{(p)}(\omega_0,\omega)= X^{(p)}(\omega_0,\omega)+X^{(p)}(\omega_0,-\omega)
\end{equation}
Unlike $X^{(p)}$, the correlator $X'^{(p)}$ is invariant upon the transformations
$\omega\rightarrow-\omega$ or $\omega_0\rightarrow-\omega_0$.
We obtain for a conductor with energy-independent transmission:
\begin{equation}
\begin{array}{rl}
X'^{(p)}(\omega_0,\omega)= & \sum_nJ_n(z)J_{n+p}(z) \left[S_2(V+n\hbar\omega_0/e,\omega) \right.\\
 & + \left.(-1)^p S_2(V-n\hbar\omega_0/e,\omega) \right]
\end{array}
\end{equation}
We notice that $X'^{(p)}$ can be expressed in terms of the noise $S_2$ and thus can be deduced from the voltage dependence of $S_2(V,\omega)$. This is not the case for $X^{(p)}$, see Eq. (\ref{eq:X}). As for $X^{(1)}$ we define the noise susceptibility $\chi'_{\omega_0}(\omega)$ associated with $X'^{(1)}$. We find:
\begin{equation}
\begin{array}{rl}
 \chi'_{\omega_0}(\omega)= & (F/2) e/(\hbar\omega_0) \left[
S_2(V+\hbar\omega_0/e,\omega) \right.\\
& - \left. S_2(V-\hbar\omega_0/e,\omega) \right]
\end{array}
\label{eq:chi'}
\end{equation}
 In the limit of a slow excitation $\omega_0\rightarrow0$, we obtain, as expected: $\chi'_0(\omega)= \frac {dS_2(\omega)}{dV}$.

\begin{figure}
\includegraphics[width= 0.75\columnwidth]{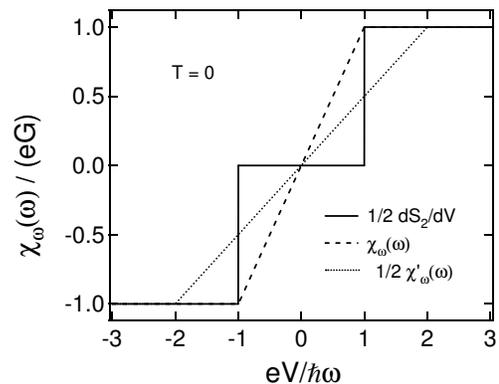}
\caption{Normalized, zero temperature noise susceptibilities $\chi_{\omega}(\omega)$ (corresponding to the setup of Fig. 1(a)) and $\chi'_{\omega}(\omega)$ (corresponding to the setup of Fig. 1(b)) vs. normalized dc bias, as well as the adiabatic response of noise given by $(1/2)dS_2(\omega)/dV$.}
\label{Fig:chi}
\end{figure}

To compare the effect of the filters in the experimental setups, we consider the noise susceptibilities $\chi$ and $\chi'$ for a conductor with energy independent transmissions, given by Eqs (\ref{eq:chi}) and (\ref{eq:chi'}). We plot them on Fig. \ref{Fig:chi} for zero temperature and in the particular case $\omega=\omega_0$, which is the one that has been investigated experimentally \cite{JGBRmanip}. We also show the adiabatic response $(1/2)dS_2(\omega)/dV$ to emphasize that at high excitation frequency $\omega_0$, the response of noise is different from what could be expected from a dc measurement. In particular, for $eV<\hbar\omega$, $\chi_\omega(\omega)\neq0$ whereas $S_2(V,\omega)$ is independent of $V$. The noise responds in phase with the excitation but has a voltage (and frequency) dependence that is not the adiabatic one. $X^{(1)}(\omega_0,\omega)$ has been recently measured in a tunnel junction for $\hbar\omega\sim\hbar\omega_0\gg k_BT$ \cite{JGBRmanip}. The experimental data agree quantitatively with the present results.

\emph{Discussion ---}
In the previous sections we have derived expressions for arbitrary energy-dependent transmissions and calculated how formula simplify when $t(E)$ and $r(E)$ do not depend on $E$. These assumptions are relevant almost only for the tunnel junction and the quantum point contact. In more complicated systems, energy dependence may be important. An intermediate and interesting case is that of the diffusive wire, in which the transmissions are randomly distributed but present correlations. In the dc current, the average of the transmission gives the Drude conductance and its fluctuations give rise to the Universal Conductance Fluctuations, which disappear upon disorder averaging. The transmissions also have an energy-dependent imaginary part, related to the diffusion time of the electrons along the wire. This should show up as an imaginary part in the conductance \cite{BuAC}. However, due to the strong screening in good conductors, this does not happen. The ac conductance is frequency independent up to plasma frequency or inverse RC-like time, at least for metals. The same suppression of frequency dependence holds for noise \cite{Hekking}. Shot noise persists up to frequencies much higher than the inverse diffusion time because, crudely speaking, when one electron enters the wire, it pushes one out almost immediately.

This does not matter for the noise susceptibility, as e.g. in a diffusive wire at low frequency, where the dynamics of the noise can be understood in terms of heat conduction along the sample \cite{NTI}. In general, the noise susceptibility gives the response of charge neutral excitations and thus is unaffected by screening. From Eq. (\ref{eqgeneX}) one can calculate $\chi_{\omega_0}(\omega)$ for a diffusive wire at arbitrary frequencies. It is crucial in such a calculation to take into account correlations between transmissions / reflections at different energies on the scales of $\omega$ and $\omega_0$. Such correlations disappear for energy differences larger than the Thouless energy $E_c=\hbar/\tau_D$ with $\tau_D$ the diffusion time. Such frequency dependence had already been pointed out in the third cumulant of noise \cite{Pilgram} and in the photo-assisted noise \cite{Pistolesi}. Moreover, Eq. (\ref{eqgeneX}) shows that the phase of the transmission / reflection coefficients matters, unlike the case of photo-assisted noise.

\emph{Link with Dynamical Coulomb Blockade} --- We will now show that the noise susceptibility is a central concept in the understanding of environmental effects on quantum transport, in particular on Dynamical Coulomb Blockade (DCB). A brief description of DCB in terms of classical currents helps understand the link between noise and DCB. A sample that is biased through an impedance $Z(\omega)$ experiences voltage fluctuations due to the voltage drop across $Z$ of the current fluctuations emitted by the sample itself. If the noise emitted by the sample depends on its voltage bias, the voltage fluctuations across the sample modify the statistics of current fluctuations creating a feedback, a phenomenon usually  called  feedback of the environment. More precisely, current fluctuations at frequency $\omega$ create voltage fluctuations $-Z(\omega)i(\omega)$ at the same frequency, which modulate the current fluctuations at all frequencies $\omega'$. In the particular case $\omega'=\omega$, this mechanism causes DCB, i.e. correction to the dc current \cite{NazIngold}. The quantity which measures the modulation of noise at frequency $\omega$ by an excitation at the same frequency is the noise susceptibility $\chi_{\omega}(\omega)$ that we have calculated. Thus, we may expect that DCB can be formulated in terms of $\chi_{\omega}(\omega)$, as we demonstrate now.

We consider only DCB in two limits: i) a tunnel junction (of small transmission) connected to an arbitrary $Z(\omega)$ (Eq. (51) in \cite{NazIngold}); ii) a sample of arbitrary transmission connected to a small $Z(\omega)$ (Eq. (7) in \cite{LevyYeyati}). In these results, one recognizes easily the expressions giving our noise susceptibility for $\omega_0=\omega$, so we can rewrite the correction to the dc current as:
\begin{equation}
\delta I(V)=-\frac{\hbar}{e^2}
\int_{-\infty}^{+\infty} K(\omega)\omega\chi_{\omega}(\omega)d\omega
\label{DCB}
\end{equation}
where $K(\omega)$ is the Fourier transform of $\exp J(t)$ and $J(t)$ is the phase-phase correlation function of the environment alone \cite{NazIngold}. Case i) is obtained by taking $1$ for the Fano factor in $\chi_\omega(\omega)$. Case ii) is recovered by expanding $\exp J(t)\sim1+J(t)$. Thus our Eq. (\ref{DCB}) not only relates noise to DCB in a transparent form, it also provides a natural extension to existing results and a physical meaning to a more general result, Eq. (8.4) of ref. \cite{Kinder}.

Feedback of the environment also strongly affects the third cumulant of voltage/current fluctuations $S_3(\omega,\omega')$ \cite{Beenakker,S3BR}. As for DCB, voltage fluctuations
at $\omega$ modulate noise at $\omega'$, which contributes to modify $S_3(\omega,\omega')$. At low frequency this effect is described by $S_2(\omega)[dS_2(\omega')/dV]$. It is clear that at high enough frequencies, this result will be modified into $S_2(\omega)\chi_{\omega}(\omega')$. Thus the understanding of the noise susceptibility is crucial to future studies of higher order cumulants at finite frequency, in particular in the quantum regime.

We thank M. Aprili, M. Devoret, D. Est\`eve, G. Montambaux, F. Pierre, H. Pothier, J.-Y. Prieur, D.E. Prober, I. Safi and C. Urbina for fruitful discussions. This work was supported by ANR-05-NANO-039-02.


\end{document}